\documentclass{sig-alternate}
\usepackage{hyperref}

\begin{document}

\title{On the Improvement of Quality and Reliability of Trust Cues in Micro-task Crowdsourcing (Position paper)}

\numberofauthors{2}

\author{
% 1st. author
\alignauthor
Jie Yang\\
       \affaddr{Delft University of Technology}\\
       \affaddr{Mekelweg 4, 2628CD Delft, The Netherlands}\\
       \email{j.yang-3@tudelft.nl}
% 2nd. author
\alignauthor
Alessandro Bozzon\\
       \affaddr{Delft University of Technology}\\
       \affaddr{Mekelweg 4, 2628CD Delft, The Netherlands}\\
       \email{a.bozzon@tudelft.nl}
}

%\date{30 July 1999}
% Just remember to make sure that the TOTAL number of authors
% is the number that will appear on the first page PLUS the
% number that will appear in the \additionalauthors section.

\maketitle
\begin{abstract}
Micro-task crowdsourcing has become a successful mean to obtain high-quality data from a large crowd of diverse people. In this context, \emph{trust} between all the involved actors (i.e. requesters, workers, and platform owners) is a critical factor for acceptance and long-term success. As actors have no expectation for ``real life'' meetings, thus trust can only be attributed through computer-mediated \emph{trust cues} like workers qualifications and requester ratings. Such cues are often the result of technical or social assessments that are performed in isolation, considering only a subset of relevant properties, and with asynchronous and asymmetrical interactions. In this paper, we advocate for a new generation of micro-task crowdsourcing systems that pursue an holistic understanding of trust, by offering an open, transparent, privacy-friendly, and socially-aware view on the all the actors of a micro-task crowdsourcing environment. 
\end{abstract}

% A category with the (minimum) three required fields
%\category{H.4}{Information Systems Applications}{Miscellaneous}
%A category including the fourth, optional field follows...
%\category{D.2.8}{Software Engineering}{Metrics}[complexity measures, performance measures]

%\terms{Theory}

%\keywords{ACM proceedings, \LaTeX, text tagging}

\section{Introduction}

\emph{Trust} is commonly defined as \emph{``an attitude of positive expectation that one's vulnerabilities will not be exploited''} \cite{trust2007}. Trust is required in situations that involve multiple actors, where there is something at stake, and where there exists a certain level of risk due to the lack of detailed knowledge about the other actors. The \emph{perception} of trust is often guided by so-called trust-warranting properties \cite{Bacharach2001}, or \emph{trust cues}, that are signalled by the involved actors.  

Identifying and signalling \emph{trust cues} is a key concern for socio-technical systems that foster trust and trustworthy behaviour \cite{Riegelsberger:2005}. \emph{Trust cues} can be observable or non-observable; they come from personal relationships and face-to-face interaction (e.g. gestures and behaviour), can be related to context (e.g. time, or social embedding), or be intrinsic of the trusted actor (e.g. expertise and motivation). 

\begin{figure}
    \centering
    \includegraphics[width=0.48\textwidth]{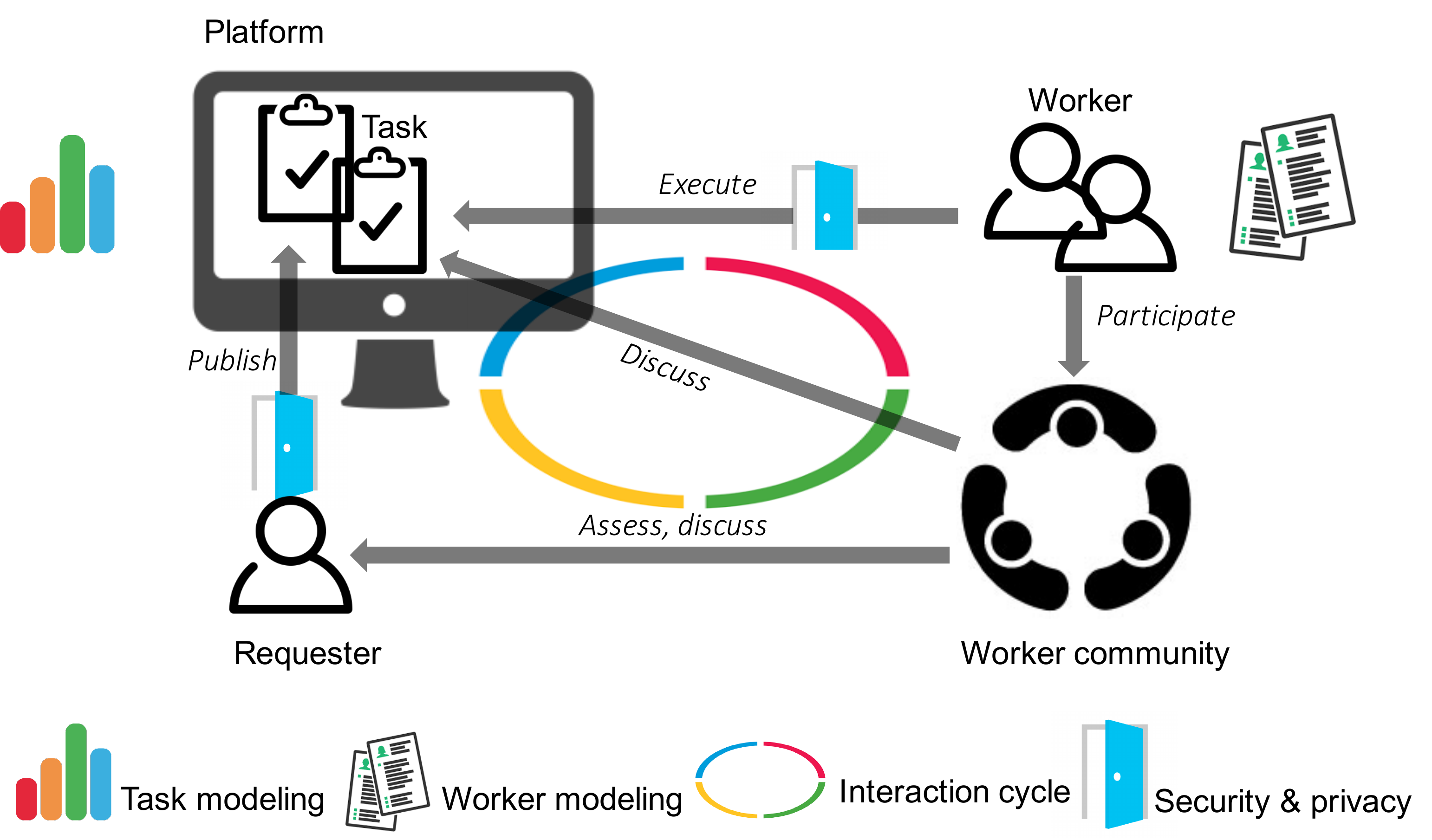}
    \caption{An overview of the trust \emph{ethos} creation process in micro-task crowdsourcing. Grey arrows represents current methods, while coloured icons symbolise the main elements of our proposal.}
    \label{fig:proposal}
    \vspace{-12pt}
\end{figure}

In the context of micro-task crowdsourcing, personal interactions are often not possible, thus resorting to computer mediated cues for mutual trust perception. These cues are typically created as depicted in Figure \ref{fig:proposal} (grey arrows).

Crowd workers are (supposedly) anonymous to the requester, yet identifiable. This is a precise design choice of the hosting platforms, including \texttt{Amazon Mechanical Turk (AMT)},  \texttt{CrowdFlower}, \texttt{Microworkers}, etc., that allows for a certain degree of privacy (at least for workers), while enabling requesters to build a perception of trust  that is extrapolated from historical knowledge about (successful) past task executions. When historical knowledge is missing, the platform provides cues -- like \texttt{AMT}'s approval rate and master qualification, or \texttt{CrowdFlower} levels -- that should help requesters with minimising the risk of low quality work. On the other hand, the identity of requesters is typically revealed; this allows workers to build, over time, a perception of trust that can guide the selection of tasks to work on, minimising the risk of unfair treatment or payment. Workers share their opinions about requesters in online community driven platforms like \texttt{Turkopticon} and \texttt{mTurkForum}. There, workers discuss about the quality and convenience of available tasks, but also about trust cues such as fairness, communication speed, and adherence to established norms\footnote{For instance \url{http://crowdsourcing-code.com}}. 

%COMMENT: MAYBE ADD SOMETHING ABOUT PLATFORMS? E.G.: https://redd.it/48rtgk

In this way, workers and requesters build an \emph{ethos} of trust that, while being critical for acceptance and long-term success of micro-task crowdsourcing, is currently based on fragmented, opaque, and often incomplete knowledge. 

\section{Issues with Trust Cues Creation}

State-of-the-systems suffer from several shortcomings that hinder the development of a reliable and sustainable trust-aware micro-task crowdsourcing. In this paper, we focus on the following four issues related to trust cues: 1) Reliance on \textbf{Result-driven} measures; 2) \textbf{Asymmetry and Fragmentation};  3) \textbf{Stagnancy}; and 4) \textbf{Asynchronicity} of interaction between the involved actors.

\vspace{5px}
\noindent\textbf{Result-centred measures}. Trust clues are currently built on a ``result-centred'' interpretation of reliability. Workers are deemed trustworthy according to their ability to successfully execute tasks, while requesters are mainly assessed according to their inclination towards acceptance and fast payment. We find this interpretation incomplete and fundamentally unfair, as it ignores relevant cues related to workers, requesters, \emph{and tasks}. For instance, the \emph{expertise} and \emph{motivations} of workers are often ignored at task evaluation time, thus hindering a fair assessment of the actual performance of the worker. Likewise, important properties of tasks such as \emph{complexity} and \emph{clarity} are not explicitly considered when assessing workers performance and requesters fairness. 

\noindent\textbf{Asymmetry and Fragmentation}. Trust cues are now produced and collected in isolation, with limited visibility for all the involved actors. For instance, cues about requesters are now exclusively exchanged in workers communities, while each requester builds its own historical, fine-grained knowledge about workers. Platform owners are not incorporating such cues as an explicit signal in the platforms, thus creating uncertainty, and an intrinsic lack of trust. 

\vspace{3px}
\noindent\textbf{Stagnancy}. Actors feature an evolutionary behaviour: they \emph{learn} new competences and skills, while building \emph{awareness} of their abilities and rights; they interact to form \emph{communities} that share norm and belief about, for instance, what constitutes fair behaviour; and they vary their involvement and \emph{availability}. These contextual and socially-aware clues are often ignored, providing a rather stagnant understanding of the trust-related properties of the involved actors. 

\vspace{3px}
\noindent\textbf{Asynchronicity}. The current paradigm of interaction between actors is asynchronous and, often focused on the \emph{outcomes} of a task execution. This prevents actors from building a shared understanding of their functional (e.g. expected outcomes) and non-functional (e.g. ROI of task learning) goals, thus ignoring once more useful contextual clues.

\vspace{-5px}
\section{Toward Richer and More Reliable Trust Cues}

Figure \ref{fig:proposal} depicts an high-level view of an hypothetical trust-aware crowdsourcing platform. We advocate for an evolution where trust cues are exchanged in an open, transparent, yet privacy-friendly manner. To build a socially shared sense of trust, we propose the adoption of a variety of methods drawn from related disciplines such as user modelling and HCI. The goal is not only to increase the amount of trust cues available in the platforms, but also to devise novel ways to support \emph{trustworthy actions} based on such cues. In the following we outline the design aspects and directions that we believe are the most crucial.

\vspace{5px}
\noindent\textbf{Task Analysis and Modelling.}
\emph{Tasks} are the objects that influence the most of the trust cues associated with performers (e.g. performance) and requester (e.g. generosity, fairness). Yet, their intrinsic properties (e.g. complexity, clarity, and usability) are not considered for assessment purposes. We advocate the need for \emph{objective} and \emph{semi-automatic} task analysis and modelling capabilities in platforms. This will allow to: 1) promptly point out most of the issues in task design, by giving requesters feedback for quality improvement based on mutually agreed guidelines; and 2) allow a better estimation of the effort required to  complete a task, thus better regulating issues related to fair payments and rewards. 

\vspace{5px}
\noindent\textbf{Open and Extensible Profiles.}
We advocate for a transparent and extensible application of advanced user modelling techniques to describe all involved actors in a more comprehensive manner, e.g. by  also in terms of capabilities, skills, motivations, and personal traits. Such novel properties could be designed and validated with the support and guidance of both requesters and workers, thus allowing for informed and, ultimately, reliable attribution. Note that this will allow the assignment of explicit roles to crowd workers, with different responsibilities and rewards, as also envisioned in previous work \cite{Demartini15}.

\vspace{5px}
\noindent\textbf{Privacy.}
Enabling advanced profiling capabilities, based on personal worker and work information, brings obvious security and privacy implications. Respecting actors, privacy is not only required from a legal and an ethical point of view, but also has critical implication for trust. However, we must stress how trust relationships are built on knowledge, thus demanding for some information to be shared. We believe privacy issues to be addressed along traditional dimensions. For instance, explicit \emph{access control} (\emph{opt-in}) can enable fine-grained access policies based on actors' profiles, thus allowing workers to decide when and with whom share their information. Indeed, real and virtual identity (or identities) shall be separated; however, workers could decide to make links between different virtual identities available (to help with modelling), or explicitly forbid any attempt to do so. 

\vspace{5px}
\noindent\textbf{The Role of Communities.} On-line worker communities are now ``isolated'' from crowdsourcing platforms, but the great amount of relevant information about work, platforms, and requesters therein produced and shared could be of great common value. Communities could also become active actors, for instance by supporting the process of task creation and improvement: future systems can include an explicit ``sandbox'' for requesters and actual workers to play with, so to allow a better alignment between the expectations of both workers and requesters \emph{before} work takes place \cite{cheng2015measuring}. By devising proper reward schemes, this will allow the creation of a participated crowd work environment, where trust and value could be created at the same time. 

%\vspace{5px}
%\noindent\textbf{Working as Learning.} Worker abilities and skills in learning process, for which working can be viewed as the source from which crowd can learn, as well as the result of learning transfer. Learning is therefore a indispensable part to foster a developing and engaging crowd work environment. ...

\vspace{-5px}
%
% The following two commands are all you need in the
% initial runs of your .tex file to
% produce the bibliography for the citations in your paper.
\small
\bibliographystyle{abbrv}
\bibliography{sigproc}  % sigproc.bib is the name of the Bibliography in this case
% You must have a proper ".bib" file
%  and remember to run:
% latex bibtex latex latex
% to resolve all references
%
% ACM needs 'a single self-contained file'!
%
%APPENDICES are optional
%\balancecolumns

%\balancecolumns % GM June 2007
% That's all folks!
\end{document}